\documentclass[journal]{IEEEtran}
\usepackage{amsmath}
\usepackage{amssymb}
\usepackage{amsthm}
\usepackage{cite}
\usepackage{color}
\usepackage{xcolor}

\usepackage{epsfig,latexsym}
\usepackage{float}
\usepackage{fancyhdr}
\usepackage{graphicx}
\usepackage{indentfirst}
\usepackage{mdwmath}
\usepackage{mdwtab}
\usepackage{subfigure}
\usepackage{setspace}
\usepackage{times}
\usepackage{url}
\usepackage{multirow}
\usepackage{algorithm}
\usepackage{algorithmic}
\usepackage{verbatim}
\usepackage{epstopdf}

\newtheorem{remark}{Remark}
\newtheorem{theorem}{Theorem}

\newtheorem{lemma}{Lemma}

\newtheorem{corollary}{Corollary}

\begin{document}

\title{Pinching-Antenna Systems (PASS)-Based User-Side Navigation: An Anchor-Line-based Approach}

\author{Zongyi Li, Jun Wang, Tianwei Hou,~\IEEEmembership{Member,~IEEE,}
and Anna Li,~\IEEEmembership{Member,~IEEE}


\thanks{This work was supported in part by the Beijing Natural Science Foundation L232041. (Corresponding author: Tianwei Hou) }

\thanks{Zongyi Li and Jun Wang are with the School of Electronic and Information Engineering, Beijing Jiaotong University, Beijing 100044, China (email: 22211385@bjtu.edu.cn; wangjun1@bjtu.edu.cn).}
\thanks{Tianwei Hou is with the Beijing Key Laboratory of Convergent Communications and Networking Technologies in LEO Satellite Systems, and also with the State Key Laboratory of Networking and Switching Technology, Beijing University of Posts and Telecommunications, Beijing 100876, China (email: htw@bupt.edu.cn).}
\thanks{Anna Li is with the School of Computing and Communications, Lancaster University, Lancaster LA1 4WA, U.K. (e-mail: a.li16@lancaster.ac.uk).}

}

\maketitle

\begin{abstract}
Pinching-antenna systems (PASS) are capable of dynamically reconfiguring wireless channels by flexibly repositioning pinching antennas (PAs) along the waveguides to establish short-range line-of-sight links. 
In this paper, a user-side navigation framework for PASS is proposed, where mobile users determine their own positions using only downlink broadcast signals without any prior knowledge of the PA positions. 
First, a Lambert W function-based PA positioning and pseudorange estimation (LWF-PAP) algorithm is developed, in which 
the closed-form expressions for both the PA positions along the waveguide and the PA-user pseudoranges are derived. 
Second, a weighted least squares-based PASS navigation (WLS-PAN) algorithm is formulated, where the nonlinear PASS-based navigation equations are transformed into a closed-form linear system,
and the optimal weight matrix is derived, 
achieving minimum-variance unbiased estimation. 
Third, the PA-derived position dilution of precision (PA-PDOP) metric is further defined to characterize the theoretical accuracy bound. 
Simulation results demonstrate that centimeter-level positioning accuracy is achieved for both PAs and users within the breakpoint distance. It is also shown that uniform PA deployment and a moderate increase in the number of PAs effectively improve navigation accuracy, 
thereby validating the effectiveness and robustness of the proposed framework for distributed real-time user-side self-navigation.
\end{abstract}

\begin{IEEEkeywords}
Lambert W function, navigation, PASS, positioning, weighted least squares.
\end{IEEEkeywords}

\section{Introduction}
The rapid evolution of next-generation wireless networks has created an unprecedented demand for high-precision positioning and navigation \cite{Wang2024}. 
In the context of the sixth-generation (6G) wireless communication systems, integrated sensing and communication (ISAC) has emerged as a key trend, offering new opportunities for high-precision positioning~\cite{Liu2022}. 
However, achieving high-accuracy positioning with ISAC requires antenna systems with flexible beam control, large-scale spatial diversity, and the ability to exploit near-field effects~\cite{liu2021reconfigurable, hou2026transmission}.
Consequently, flexible antenna systems, such as movable antennas~\cite{Zhu2024, Zhu2024_MA_Modeling}, fluid antennas~\cite{Huangfu2026_FAS_Rician, Wong2020b}, and pinching-antenna systems (PASS)~\cite{Ding2024, Liu2026}, have attracted research interest.

Among the emerging flexible antenna systems, PASS stands out as a breakthrough paradigm for large‑scale reconfigurable wireless architectures~\cite{hou2026performance}.
PASS is a system of one or more dielectric waveguides along which multiple pinching antennas (PAs) are deployed at dynamically tunable positions.
Signals propagate along the waveguide with low attenuation before radiating into free space at designated pinch points~\cite{Ding2024}. 
The key idea of PASS is to create strong line-of-sight (LoS) links by using the waveguide to flexibly position PAs close to users, bringing wireless communications from the ``last mile'' to the ``last meter''~\cite{Liu2026}. 
By leveraging the waveguide, PASS can achieve meter‑scale antenna reconfiguration~\cite{Ouyang2025, Hou2026}, which is in stark contrast to most other reconfigurable schemes that are typically confined to altering channel characteristics within the wavelength scale~\cite{Zhu2024}. 
Therefore, PASS distinguishes itself by facilitating large-scale antenna reconfiguration, scalable implementation, and near-field benefits. 


The unique characteristics of PASS have motivated a growing body of research exploring its potential for positioning and sensing applications. 
For the AP-side user positioning, several studies have been conducted. 
In~\cite{Zhang2025}, a PASS-based received signal strength indication (RSSI) method for distance measurement was proposed, demonstrating the feasibility of indoor positioning using PASS. 
In~\cite{Xu2025}, a PASS-based joint user positioning and channel estimation framework was developed, where PAs are grouped into subarrays to cooperatively estimate user and scatterer positions. 
In~\cite{Liu2025_OFDM}, a PASS-based joint positioning and channel estimation framework with orthogonal frequency-division multiplexing (OFDM) was established, and the Cramér-Rao lower bound (CRLB) for user position estimates was derived, characterizing the fundamental limits of AP-side user positioning accuracy.
For the user-side self-positioning, the research is still in an early stage. In~\cite{He2026}, a stochastic geometry framework for PASS-based user-side self-positioning was established, 
and a closed-form expression of the CRLB distributions was derived, characterizing the fundamental limits of AP-side user positioning accuracy.
Collectively, recent works have established that the PASS is a promising enabler for high-accuracy positioning. 

It is notable that the existing PASS-based positioning studies commonly operate under a particular assumption: the positions of PAs along the waveguide are considered known~\cite{Zhang2025, Xu2025, Liu2025_OFDM, He2026}. With this assumption, each received signal can be associated with a known PA position, and the positioning problem is thereby converted into a conventional anchor-node scenario. 
Moreover, in the AP-side positioning frameworks, user positioning is performed centrally at the AP, and users are not involved in computing their own positions. 
While such AP-side processing is suitable for applications where centralized positioning is sufficient, it cannot satisfy the demands of user-side navigation. 

In user-side navigation scenarios, mobile users estimate their positions in real time by using downlink signals, without relying on continuous uplink signaling or centralized computation~\cite{hou2023performance}. 
The global navigation satellite systems (GNSS) represents the most mature implementation. In non-denied environments, users can triangulate their positions using time-of-arrival (TOA) measurements from multiple GNSS satellites with precisely known ephemeris~\cite{Misra2011}. However, in denied environments, GNSS fails to provide reliable positioning due to severe signal attenuation, multipath propagation, and the absence of LoS links to satellites \cite{Ghafoori2025ImprovedGNSS}. 
Therefore, other user-side navigation systems were explored for denied environments. 
Ultra-wideband (UWB) systems broadcast signals from anchor nodes with known coordinates, allowing users to estimate their own positions via two-way ranging or time-difference-of-arrival (TDoA) techniques~\cite{Gezici2005, Neirynck2016}. 
Wi-Fi fingerprinting constructs location-dependent signal maps from multiple access points, enabling user-side navigation through pattern matching~\cite{gao2025loc, Faragher2015}. 
Bluetooth low-energy (BLE) beacons with known deployment positions enable users to compute their coordinates based on proximity and signal strength measurements~\cite{Pau2021, Kwon2025}.
However, conventional user-side navigation approaches suffer from fixed and limited coverage, multipath-induced accuracy degradation, and sensitivity to indoor environmental changes~\cite{Yang2023}.
Moreover, all of the approaches depend on precisely located anchor nodes with known coordinates, adding the overhead of deployment and transmission. 

\subsection{Motivation and Contributions}


From the above discussion, it is clear that PASS-based user-side navigation is a promising but largely unexplored direction. 
Although a preliminary theoretical study for PASS-based user-side positioning has been reported in \cite{He2026}, it still operates under the assumption that the PA positions are perfectly known to the users. 
In other words, the PAs are treated as conventional anchor nodes with precisely calibrated coordinates. 
The assumption essentially strips PASS of its most compelling advantage: the ability to dynamically and flexibly reposition the PAs without the burden of manual calibration. 
In PASS, PAs may be frequently reconfigured to track users or adapt to environmental changes. 
Therefore, for practical user‑side navigation, requiring a mobile terminal to always know the exact, up‑to‑date PA positions is highly unrealistic. 
Continuously broadcasting PA coordinates not only consumes precious downlink resources but also conflicts with the privacy and autonomy that characterize user‑side navigation.


The main motivation for this work is that realizing PASS-based user-side navigation without the precise PA positions would unlock the true “drop‑and‑position” capability of PASS, where user‑side navigation becomes available immediately after deployment, without any site‑specific calibration. 
However, three fundamental challenges arise:


\begin{itemize}
    \item 
    Unknown PA positions at user side: 
    In conventional anchor-node-based navigation, users obtain reference point coordinates via broadcast ephemeris or pre-loaded maps, which is feasible because that reference nodes have predictable or static positions. 
    In PASS, the AP dynamically activates PAs along the waveguide, and the activated PA positions are neither predetermined nor fixed across time slots. Consequently, users cannot predict or acquire the PA coordinates a priori, which renders conventional anchor-node-based methods inapplicable.
    
    \item 
    Coupled  propagation components:
    The received downlink signal in PASS aggregates two distance-dependent components, namely the in-waveguide propagation loss which depends on the PA-AP distance, and the free-space path loss which depends on the PA-user distance. Since PA positions are unknown to the user, these two components cannot be directly separated. 
    Likewise, the measured propagation time also contains both the in-waveguide delay and the free-space delay, preventing any conventional one-to-one mapping between propagation time and distance.

    \item 
    Inapplicability of conventional trilateration methods:
    Classical trilateration inherently relies on anchor nodes spanning at least a two-dimensional spatial configuration. 
    However, in PASS, a single waveguide confines all PAs to a one-dimensional straight line. 
    As a result, any set of distance measurements from the PAs yields a locus of possible user locations that is rotationally symmetric about the waveguide axis. 
    Even if the PA positions were known, the geometric degeneracy would prevent a unique determination of the user coordinates, rendering conventional trilateration methods inapplicable.
\end{itemize}

These intertwined challenges have not been addressed in any prior work on PASS-based positioning. Thus, this paper proposes a novel PASS-based user-side navigation approach that empowers mobile users to determine their own positions using only downlink broadcast signals, without requiring prior knowledge of PA positions and without relying on AP-side processing for position computation. The main contributions of this paper are summarized as follows.

\begin{itemize}
    \item 
    We propose a PASS-based user-side navigation framework, in which an AP broadcasts downlink signals through multiple PAs along a single waveguide. Each user independently estimates its own position using only the received downlink measurements, without requiring any prior knowledge of the PA positions. The framework achieves distributed self-positioning on the user side, enabling real-time positioning with minimal signaling overhead and without reliance on centralized computation.
    
    \item 
    To address the dual challenges of unknown PA positions and coupled propagation components, we propose a Lambert W function-based PA positioning and pseudorange estimation (LWF-PAP) algorithm. By leveraging two heterogeneous measurements, namely received signal power characterizing path loss and time of arrival characterizing propagation delay, we transform the coupled nonlinear system into a tractable form, obtaining closed-form expressions for both the PA position along the waveguide and the PA-user pseudorange. We establish the necessary and sufficient condition for the existence and uniqueness of the solution, and analyze the error propagation characteristics. 
    
    \item 
    Building upon the estimated PA positions and pseudoranges from multiple PAs, we further develop a weighted least squares-based PASS navigation (WLS-PAN) algorithm that optimally fuses measurements to compute the user position. The nonlinear navigation equations are transformed into a closed-form linear system through a novel auxiliary variable formulation, where the inherent solution ambiguity is resolved by using corridor boundary constraints.
    Based on the Gauss–Markov theorem, we derive the optimal weight matrix, enabling minimum-variance unbiased estimation. 
    We also define the PA-derived position dilution of precision (PA-PDOP) metric to general evaluate the performance of the proposed anchor-line-based approach for PASS-based user-side navigation.
    
    \item 
    The simulation results confirm our analysis and also demonstrate that: i) The proposed LWF-PAP algorithm achieves centimeter-level PA positioning accuracy within the breakpoint distance; ii) The proposed anchor-line-based approach for the PASS-based user-side navigation enables centimeter-level user navigation accuracy within the working range; iii) The proposed WLS-PAN algorithm achieves a theoretical PA-PDOP bound below 0.3, indicating a high precision upper limit. 
    Moreover, uniform PA deployment along the waveguide is shown to improve both accuracy and robustness compared to random deployment. In addition, increasing the number of PAs effectively reduces user navigation error, though the benefit gradually diminishes. These results collectively validate the superiority of our proposed framework.
\end{itemize}

\subsection{Organization}

The remainder of this paper is organized as follows. Section II presents the model of the proposed PASS-based navigation system.
Section III details the proposed navigation approach, comprising the LWF-PAP algorithm with error analysis, the WLS-PAN algorithm, and the PA-PDOP metric. 
Section IV provides comprehensive simulation results to evaluate the performance of the proposed algorithms under various system parameters.
Finally, Section V concludes the paper with a summary of key findings and directions for future research.

\section{System Model}

\begin{figure}[t!]
\centering
\includegraphics[width =3.4in]{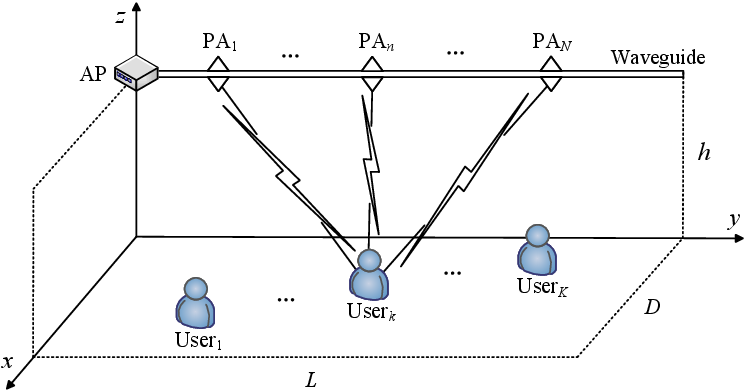}
\caption{PASS-based navigation system in a long corridor with a single multi-PA waveguide.}
\label{system_model}
\end{figure}

Fig.~\ref{system_model} illustrates the considered PASS-based navigation system, where an AP broadcasts downlink navigation signals to $K$ users through a single waveguide with $N$ PAs. 
The waveguide is deployed along the ceiling boundary of a long corridor, which has a length of $L$ and a width of $D$. 
To satisfy the user-side navigation requirement, the number of PAs must be greater than the plane dimension, i.e., $N\geq2$. 
The geometry of the waveguide is described by a line segment $\mathbf{G}_n = [0, (0,L), h]^{\mathrm{T}}$. 

The position of the AP is given by $\mathbf{w}_0 = [0, 0, h]^{\mathrm{T}}$. The position of the $n$-th PA is expressed as $\mathbf{w}_n = [0, y_{n}, h]^{\mathrm{T}}$, for $n \in \{ 1,2,\dots,N \}$, 
where $y_n \in (0,L)$ is the distance between the AP and the $n$-th PA along the waveguide. 
For the users, it is noted that the positions of PAs are unknown, i.e., $y_n$ is unknown. 
The position of the $k$-th user is assumed to be $\mathbf{u}_k = [x_{\mathrm{u},k}, y_{\mathrm{u},k}, 0]^{\mathrm{T}}$, for $k \in\{1,2,\ldots,K\}$, where $x_{\mathrm{u},k} \in (0,D)$ and $y_{\mathrm{u},k} \in (0,L)$.

In this paper, we adopt an ideal waveguide model, ignoring the performance degradation induced by the volume and shape of the waveguides. 
We assume full coupling between PAs and waveguides, corresponding to full-power radiation of the PAs.

\subsection{Channel Model}
We first characterize the free-space channel between the PAs and the users under a LoS assumption~\cite{Mu2025_Multicast, Xu2026_PASS_LoS_Blockage}. The large-scale fading coefficient between the $n$-th PA and the $k$-th user is modeled as 
\begin{equation}\label{hls}
l_{nk} = {\frac{\eta}{d_{nk}}},
\end{equation}
where $\eta=\frac{c}{4 \pi f_c}$, $c$ is the speed of light in free space, $f_c$ is the carrier frequency, while $d_{nk}$ is the distance between the $n$-th PA and the $k$-th user, which is given by
\begin{equation}\label{dfs}
\begin{split}
d_{nk} &= \|\mathbf{w}_n - \mathbf{u}_k\|\\
&= \sqrt{x_{\mathrm{u},k}^2 + {\left( y_{n} - y_{\mathrm{u},k} \right)}^2 + { h }^2},
\end{split}
\end{equation}
where $\|\cdot\|$ denotes the Euclidean norm. 

The propagation delay of the signal over the free-space channel from the $n$-th PA to the $k$-th user is defined as
\begin{equation}\label{tau_fs}
\tau_{nk} = \frac{d_{nk}}{c},
\end{equation}
which directly determines the phase rotation of the signal in the free-space channel. 
By assuming that the channel gain is normalized to unity, the small-scale channel can be modeled as 
\begin{equation}\label{hss}
\begin{split}
m_{nk}&= \exp\left(-j 2 \pi f_c \tau_{\mathrm{fs},nk}\right)\\
&= \exp\left(-\frac{j 2 \pi d_{nk}}{\lambda}\right),
\end{split}
\end{equation}
where $\lambda$ is the carrier wavelength. 

Thus, the gain of the free-space channel between the $n$-th PA and the $k$-th user is modeled as 
\begin{equation}\label{hfs}
\begin{split}
h_{nk} &= l_{nk} m_{nk} \\
&= {\frac{\eta}{d_{nk} }}\exp\left(-\frac{j 2 \pi d_{nk}}{\lambda}\right).
\end{split}
\end{equation}

For the in-waveguide channel from the AP to the $n$-th PA, the channel gain is modeled as \cite{Xu2025_PASS_Attenuation}
\begin{equation}\label{hiw}
g_{n} = e^{-(\alpha + j\beta) y_{n} },
\end{equation}
where $\alpha= \frac{\pi \tan \delta}{\lambda_\mathrm{g}}$ and $\beta = \frac{2 \pi}{\lambda_\mathrm{g}}$ are the attenuation constant and phase constant of the waveguide medium, respectively. $\tan \delta$ is the waveguide dielectric dissipation factor, ${\lambda_\mathrm{g}}=\frac{\lambda}{\sqrt{\varepsilon_\mathrm{r}}}$ denotes the waveguide wavelength, where $\varepsilon_\mathrm{r}$ is the relative permittivity of the waveguide medium. 

Correspondingly, the propagation delay of the signal over the in-waveguide link from the AP to the $n$-th PA is defined as \cite{Pozar2012}
\begin{equation}\label{tau_iw}
\tau_{\mathrm{g},n} = \frac{y_n}{v_\mathrm{g}},
\end{equation}
where $v_\mathrm{g} = \frac{c}{\sqrt{\varepsilon_\mathrm{r}}}$ denotes the group velocity of the electromagnetic wave in the waveguide medium. 

Therefore, the total propagation delay from the AP to the $k$-th user via the $n$-th PA is the sum of the in-waveguide delay and the free-space delay, given by
\begin{equation}\label{T_total}
\begin{split}
\bar{T}_{nk} &= \tau_{\mathrm{g},n} + \tau_{nk} \\
&= \frac{1}{c}\left({\sqrt{\varepsilon_\mathrm{r}}y_n + d_{nk}}\right).
\end{split}
\end{equation}

Based on the above, the effective channel gain from the AP to the $k$-th user via the $n$-th PA is expressed as 
\begin{equation}\label{channel}
\begin{split}
\bar{h}_{nk} &= g_{n} h_{nk}\\
&=\frac{\eta e^{-\alpha y_{n}}}{d_{nk}} \exp\left(-j 2 \pi\left(\frac{ d_{nk}}{\lambda}+\frac{ y_{n}}{\lambda_\mathrm{g}}\right)\right).
\end{split}
\end{equation}

\subsection{Signal Model}
In the downlink transmission, we consider time division multiple access (TDMA) for allocating wireless resources to different PAs. The total available time resource block is divided into $N$ time slots, and the $n$-th time slot is exclusively assigned to the $n$-th PA. 
Thus, the signal received by the $k$-th user in the $n$-th time slot, namely the $n$-th signal, can be expressed as 
\begin{equation}\label{received signal}
\hat{r}_{nk} = \bar{h}_{nk} \sqrt{P_{n}} s_n + n_k,
\end{equation}
where $P_{n}$ represents the downlink transmit power injected by the AP during the $n$-th time slot. $s_n$ denotes the symbol transmitted by the $n$-th PA, with $\mathbb{E}\left(|s_n|^2\right)=1$ where $\mathbb{E}\left(\cdot\right)$ denoting the expectation operation. $n_k \sim \mathcal{CN}(0, \sigma^2)$ represents the additive white Gaussian noise (AWGN) at the $k$-th user.

For the TOA measurement, we employ bidirectional time synchronization to eliminate the clock offset between the AP and the users \cite{Neirynck2016}.
Thus, the reception time of the $n$-th signal at the $k$-th user can be given by
\begin{equation}\label{rec time}
\hat{t}_{nk} = t_n + \bar{T}_{nk} + \tilde{T}_{nk},
\end{equation}
where $t_n$ is the transmission timestamp from the AP, and $\tilde{T}_{nk}\sim\mathcal{N}(0,\sigma_{\mathrm{t},nk}^2)$ denotes the TOA measurement error of the $n$-th signal at the $k$-th user \cite{Gezici2005}.

\section{PASS-Based Navigation Approach}
In this section, we first describe the geometric positioning principle. 
Compared with traditional anchor-node-based navigation methods, 
we propose a novel anchor-line-based navigation approach for PASS. However, one natural question arises: the positions of PAs are unknown to users, where an additional dimension is evolved. 
Thus, we further propose a two-step PASS-based navigation approach, including LWF-based PA positioning and pseudorange estimation, and weighted least squares method (WLS)-based user-side navigation.

\subsection{LWF-based PA Positioning and Pseudorange Estimation}
Unlike the conventional anchor-node-based navigation, PASS-based navigation needs to jointly recover one in-waveguide pseudorange and one free-space pseudorange. To estimate these two distance terms, two heterogeneous measurements are required as independent observation constraints. In this paper, we adopt the path loss derived from received signal power and the propagation time obtained from the TOA as the two heterogeneous measurements.

Based on~\eqref{channel}, the path loss from the AP to the $k$-th user via the $n$-th PA can be given by
\begin{equation}\label{Lnk}
\bar{L}_{nk} = \frac{d_{nk}^2}{\eta^2}e^{2\alpha y_n}.
\end{equation}
In decibel scale,~\eqref{Lnk} is rewritten as
\begin{equation}\label{Lnkdb}
\bar{L}_{nk}|_{\mathrm{dB}} = 20\lg\left(d_{nk}\right) + \frac{20}{\ln 10}\alpha y_n - 20\lg\left(\eta\right).
\end{equation}

By using~\eqref{T_total} and~\eqref{Lnkdb}, we have
\begin{equation}\label{system equation1}
\left\{
\begin{array}{l}
d_{nk}+\sqrt{\varepsilon_\mathrm{r}}y_n=c\bar{T}_{nk},\\
\ln(d_{nk})+\alpha y_n=\frac{\ln10}{20}\bar{L}_{nk}|_{\mathrm{dB}}+\ln(\eta).
\end{array}
\right.
\end{equation}

Note that \eqref{system equation1} is a coupled nonlinear transcendental system with two unknown variables $d_{nk}$ and $y_n$, which admits no unique closed-form analytical solution via conventional linear solving methods.

To guarantee the existence and uniqueness of the solution, we first give the necessary and sufficient condition in the following lemma.

\begin{lemma}\label{lemmacondition}
In PASS-based navigation systems, when PAs serve users in LoS channels, 
the proposed LWF-based PA positioning and pseudorange estimation method is valid if and only if the distance between the $n$-th PA and the $k$-th user $d_{nk}$ satisfies 
\begin{equation}\label{condition}
d_{nk}\le d_0, 
\end{equation}
where $d_0$ is the breakpoint distance, given by 
\begin{equation}\label{d0}
d_0 = \frac{\lambda}{\pi\tan\delta}.
\end{equation}
\begin{proof}
Please refer to Appendix A.
\end{proof}
\end{lemma}
\begin{remark}\label{remarklemma1}
Lemma~\ref{lemmacondition} reveals the operating boundary of LWF in PASS-based navigation. 
From~\eqref{d0}, the breakpoint distance $d_0$ is proportional to carrier wavelength $\lambda$ and inversely proportional to the waveguide dielectric dissipation factor $\tan\delta$. Therefore, systems with longer carrier wavelength and lower-loss waveguides achieve a wider operating range. 
\end{remark}

Under the validity condition given in Lemma~\ref{lemmacondition}, the system equation \eqref{system equation1} has a unique closed-form solution. Specifically, the distance between the PA-user $d_{nk}$ and the position of the waveguide of the $n$-th PA $y_n$ can be explicitly expressed by the path loss $\bar{L}_{nk}|_\mathrm{dB}$ and the propagation time $\bar{T}_{nk}$, which are derived in the following theorems.

\begin{theorem}\label{theorem1}
In the PASS, when the $n$-th PA serve the $k$-th user in a LoS channel and the distance between them is within the breakpoint distance $d_0$, 
the distance between the $n$-th PA and the $k$-th user can be expressed as
\begin{equation}\label{dnkbar}
{d}_{nk}=-d_0W_0\left(-\frac{e^{\bar{\xi}_{nk}}}{d_0}\right),
\end{equation}
where $W_0(\cdot)$ represents the principal branch of the LWF \cite{Corless1996_LambertW}, and $\bar{\xi}_{nk}=\frac{\ln10}{20}{L}_{nk}|_{\mathrm{dB}}-\frac{c}{d_0}{T}_{nk}+\ln(\eta)$.
\begin{proof}
By substituting \eqref{condition} into \eqref{eq:appendix_A6}, we can obtain \eqref{dnkbar}, and the proof is complete.
\end{proof}
\end{theorem}
\begin{theorem}\label{theorem2}
In the PASS, when the $n$-th PA serve the $k$-th user in a LoS channel and the distance between them is within the breakpoint distance $d_0$, 
the distance between the AP and the $n$-th PA can be expressed as 
\begin{equation}\label{ynkbar}
{y}_{n}=\frac{1}{\sqrt{\varepsilon_\mathrm{r}}}\left(c\bar{T}_{nk}+d_0W_0\left(-\frac{e^{\bar{\xi}_{nk}}}{d_0}\right)\right).
\end{equation}
\begin{proof}
By substituting~\eqref{dnkbar} into~\eqref{T_total}, we can obtain~\eqref{ynkbar}, and the proof is complete. 
\end{proof}
\end{theorem}

\begin{remark}\label{domain}
The physical constraints of PASS guarantee that the argument of the LWF principal branch $W_0(\cdot)$ strictly falls within its valid domain $\left[-\frac{1}{e}, 0\right)$. The valid argument range ensures a unique real-valued solution for both $d_{nk}$ and $y_{n}$. 
\end{remark}

During the $n$-th time slot, the AP broadcasts a timestamp $t_{n}$ and the transmission power $P_{n}$ to the users by the $n$-th PA, 
while the $k$-th user measures the TOA $\hat{t}_{nk}$ and received power $\hat{P}_{nk}$. Then, at the $k$-th user, the estimated propagation time $\hat{T}_{nk}$ and the estimated pass loss $\hat{L}_{nk}$ from the AP to the $k$-th user via the $n$-th PA can be calculated by 
\begin{equation}\label{estimate T}
\hat{T}_{nk} = \hat{t}_{nk}-t_{n},
\end{equation}
\begin{equation}\label{estimate L}
\hat{L}_{nk}|_{\mathrm{dB}} = 10\lg{P_{n}} - 10\lg{\hat{P}_{nk}},
\end{equation}
respectively.
The received power at the $k$-th user in the $n$-th slot can be modeled as
\begin{equation}\label{bar Pnk}
\hat{P}_{nk} = \bar{P}_{nk} + \tilde{P}_{nk},
\end{equation}
where $\bar{P}_{nk}=\bar{L}_{nk} P_n$ is the signal power at the $k$-th user, and $\tilde{P}_{nk} \sim \mathcal{N}(0, \sigma_{\mathrm{p},nk}^2)$ the impact of noise on the received power \cite{Urkowitz1967}. 


Based on~\eqref{dnkbar} and~\eqref{ynkbar}, the $k$-th user can estimate its pseudorange to the $n$-th PA and the position of the $n$-th PA on the waveguide, given by 
\begin{equation}
\begin{split}
\hat{d}_{nk} = -d_\mathrm{0} W_0\left( \hat\zeta_{nk} \right),
\end{split}
\label{dnkhat}
\end{equation}
\begin{equation}
\begin{split}
\hat{y}_{nk} = \frac{1}{\sqrt{\varepsilon_\mathrm{r}}} \left( c \hat{T}_{nk} + d_\mathrm{0} W_0\left( \hat\zeta_{nk} \right)\right), 
\end{split}
\label{ynkhat}
\end{equation}
respectively, where $\hat\zeta_{nk}$ is the estimated Lambert argument. 

According to Remark~\ref{domain}, the valid range of the Lambert argument is $\left[-\frac{1}{e}, 0\right)$. 
However, in PASS-based navigation approach, the presence of noise and TOA measurement errors may drive the Lambert argument to fall outside the effective domain, thereby resulting in a non-existent or invalid solution. 
To avoid such numerical instability and ensure a meaningful solution, we impose a range constraint on the Lambert argument $\hat\zeta_{nk}$ by clamping it to the valid domain. Accordingly, $\hat\zeta_{nk}$ is constructed as \cite{Deotti2024_LambertW, hou2020reconfigurable}
\begin{equation}
\hat\zeta_{nk} =
\begin{cases}
-\dfrac{1}{e}, & -\dfrac{e^{\hat{\xi}_{nk}}}{d_0} < -\dfrac{1}{e}, \\
-\dfrac{e^{\hat{\xi}_{nk}}}{d_0}, & -\dfrac{1}{e} \le -\dfrac{e^{\hat{\xi}_{nk}}}{d_0} \leq 0, \\
0, & -\dfrac{e^{\hat{\xi}_{nk}}}{d_0} > 0,
\end{cases}
\end{equation}
where $\hat\xi_{nk} = \frac{\ln 10}{20} \hat{L}_{nk} |_{\mathrm{dB}} - \frac{c} {d_0}\hat{T}_{nk} + \ln(\eta)$. 

Based on~\eqref{dnkhat} and~\eqref{ynkhat}, the $k$-th user can obtain the estimated PA positions $\hat{\mathbf{w}}_{nk}=[0,\hat{y}_{nk},h]^{\mathrm{T}}$ and the estimated pseudorange to the PAs. 
The corresponding procedure is summarized in \textbf{Algorithm~\ref{Algorithm 1}}.
\begin{algorithm}[H]
\caption{LWF-PAP Algorithm}\label{Algorithm 1}
\begin{algorithmic}[1]
\REQUIRE
The waveguide geometry $\mathbf{G}_n$, transmission timestamp $t_n$, transmission power $P_n$, measured TOA $\hat{t}_{nk}$, measured received power $\hat{P}_{nk}$, path-loss parameter $\eta$, breakpoint distance $d_0$, and waveguide relative permittivity ${\varepsilon_{\mathrm{r}}}$. 
\ENSURE
The estimated PA position $\hat{\mathbf{w}}_{nk}=[0,\hat{y}_{nk},h]^{\mathrm{T}}$ and the estimated pseudorange to PA $\hat{d}_{nk}$.
\STATE Compute the path loss $\hat{L}_{nk}|_{\mathrm{dB}}=10\lg(P_n)-10\lg(\hat{P}_{nk})$.
\STATE Compute the propagation time $\hat{T}_{nk}=\hat{t}_{nk}-t_n$.
\STATE Define $\hat\xi_{nk} = \frac{\ln 10}{20} \hat{L}_{nk} |_{\mathrm{dB}} - \frac{c} {d_0}\hat{T}_{nk} + \ln(\eta)$.
\STATE Define $\hat\zeta_{nk} = \mathrm{clamp} \left( \frac{1}{e} ,-\frac{e^{\hat{\xi}_{nk}}}{d_0},0 \right)$.
\STATE Compute the estimated PA position: \\
$\hat{y}_{nk} = \frac{1}{\sqrt{\varepsilon_\mathrm{r}}} \left( c \hat{T}_{nk} + d_\mathrm{0} W_0\left( \hat\zeta_{nk} \right)\right)$.
\STATE Compute the estimated pseudorange to the PA: \\
$\hat{d}_{nk} = -d_\mathrm{0} W_0\left( \hat\zeta_{nk}\right)$.

\STATE \textbf{return} $\hat{\mathbf{w}}_{nk}=[0,\hat{y}_{nk},h]^{\mathrm{T}}$ and $\hat{d}_{nk}$. 
\end{algorithmic}
\end{algorithm}

\subsection{Error Analysis}
In this subsection, we characterize the error propagation from raw measurements to PA position and pseudorange estimates, which lays the theoretical foundation for optimal weight design in the subsequent WLS-based navigation stage.

The Lambert argument $\hat{\zeta}_{nk}$ is a function of the raw measurements $\hat{T}_{nk}$ and $\hat{P}_{nk}$, whose estimation error is denoted as $\tilde {\zeta}_{nk} = \hat{\zeta}_{nk} - \bar{\zeta}_{nk}$, with $\bar{\zeta}_{nk} = \exp({\bar{\xi}_{nk}})/d_0$ denoting the true value of the Lambert argument.

Based on the first-order multivariate error propagation law, the variance of $\tilde {\zeta}_{nk}$ is derived as \cite{Kay1993_Estimation}
\begin{equation}
\begin{split}
\mathrm{Var}(\tilde {\zeta}_{nk}) 
&= \left( \frac{\partial \zeta_{nk}}{\partial T_{nk}} \right)^2 \sigma_{\mathrm{t},nk}^2 + \left( \frac{\partial \zeta_{nk}}{\partial P_{nk}} \right)^2 \sigma_{\mathrm{p}}^2 \\
&= \frac{c^2 {\zeta}^2_{nk}}{d_0^2} \sigma_{\mathrm{t},nk}^2 + \frac{{\zeta}^2_{nk}}{4 {P}^2_{nk}} \sigma_{\mathrm{p}}^2.
\end{split}
\label{eq:zeta_variance}
\end{equation}


The principal branch of the LWF exhibits strong nonlinearity, especially near its branch point $-1/e$, which leads to the amplification of the Lambert argument error in pseudorange and PA position estimation.
To quantify the nonlinear error amplification effect, we define the Lambert sensitivity factor as
\begin{equation}
\begin{split}
S_{nk} &= \left| \frac{\mathrm{d}W_0(z)}{\mathrm{d}z} \bigg|_{z={\zeta}_{nk}} \right| \\
&= \frac{\left| W_0({\zeta}_{nk}) \right|}{\left| {\zeta}_{nk} \left( 1 + W_0({\zeta}_{nk}) \right) \right| + \varepsilon},
\end{split}
\label{eq:lambert_sensitivity}
\end{equation}
where $\varepsilon$ is a small positive regularization term to avoid numerical singularity when ${\zeta}_{nk}$ approaches $-1/e$. 
Based on the first-order Taylor expansion of the LWF, the variance of the pseudorange estimation error $\tilde{d}_{nk} = \hat{d}_{nk} - d_{nk}$ is derived as
\begin{equation}
\begin{split}
\sigma_{\mathrm{d},nk}^2 &= \mathrm{Var}(\tilde{d}_{nk}) \\
&= \left( \frac{\mathrm{d}d_{nk}}{\mathrm{d}\zeta_{nk}} \right)^2 \mathrm{Var}(\tilde {\zeta}_{nk}) \\
&= c^2 S_{nk}^2 {\zeta}^2_{nk} \sigma_{\mathrm{t},nk}^2 + \frac{d_0^2 S_{nk}^2 {\zeta}^2_{nk}}{4 {P}^2_{nk}} \sigma_{\mathrm{p}}^2.
\end{split}
\label{eq:dnk_variance}
\end{equation}

Similarly, the variance of the PA position estimation error $\tilde{ y}_{nk} = \hat{y}_{nk} - y_n$ is derived as
\begin{equation}
\begin{split}
\sigma_{\mathrm{y},nk}^2 &= \mathrm{Var}(\tilde{ y}_{nk}) \\
&= \left( \frac{\mathrm{d}y_{n}}{\mathrm{d}\zeta_{nk}} \right)^2 \mathrm{Var}(\tilde {\zeta}_{nk}) \\
&= \frac{c^2(S_{nk}^2 {\zeta}^2_{nk}+1)}{\varepsilon_\mathrm{r}} \sigma_{\mathrm{t},nk}^2 + \frac{d_0^2 S_{nk}^2 {\zeta}^2_{nk}}{4 \varepsilon_\mathrm{r} {P}^2_{nk}} \sigma_{\mathrm{p}}^2.
\end{split}
\label{eq:ynk_variance}
\end{equation}


\subsection{WLS-based User-Side Navigation}
Obtaining the positions of the PAs $\hat{\mathbf{w}}_{nk}$ and the pseudorange $\hat{d}_{nk}$, we have 
\begin{equation}
\|\hat{\mathbf{w}}_{nk} - \hat{\mathbf{u}}_{k} \| = \hat{d}_{nk}, 
\end{equation}
where $\hat{\mathbf{u}}_{k} = [\hat{x}_{\mathrm{u},k}, \hat{y}_{\mathrm{u},k}, 0] $ represents the estimated position of the $k$-th user. 
According to the definition of the Euclidean norm, the $n$-th navigation equation can be expressed as  
\begin{equation}\label{navi_eq}
{\hat{x}_{\mathrm{u},k}^2 + {\left( \hat{y}_{nk} - \hat{y}_{\mathrm{u},k} \right)}^2 + h ^2} = \hat{d}^2_{nk}. 
\end{equation}

Therefore, we can establish the navigation equations as 
\begin{equation}\label{system equation2}
\begin{split}
\left\{
\begin{array}{l}
{\hat{x}_{\mathrm{u},k}^2 + {\left( \hat{y}_{1k} - \hat{y}_{\mathrm{u},k} \right)}^2 + h^2} = \hat{d}^2_{1k}  \\
{\hat{x}_{\mathrm{u},k}^2 + {\left( \hat{y}_{2k} - \hat{y}_{\mathrm{u},k} \right)}^2 + h^2} = \hat{d}^2_{2k} \\
\quad \quad \quad \quad\quad\quad\quad \vdots \\
{\hat{x}_{\mathrm{u},k}^2 + {\left( \hat{y}_{Nk} - \hat{y}_{\mathrm{u},k} \right)}^2 + h^2} = \hat{d}^2_{Nk} \\
\end{array}
\right.. 
\end{split}
\end{equation}

\begin{figure}[t!]
\centering
\includegraphics[width =3.3in]{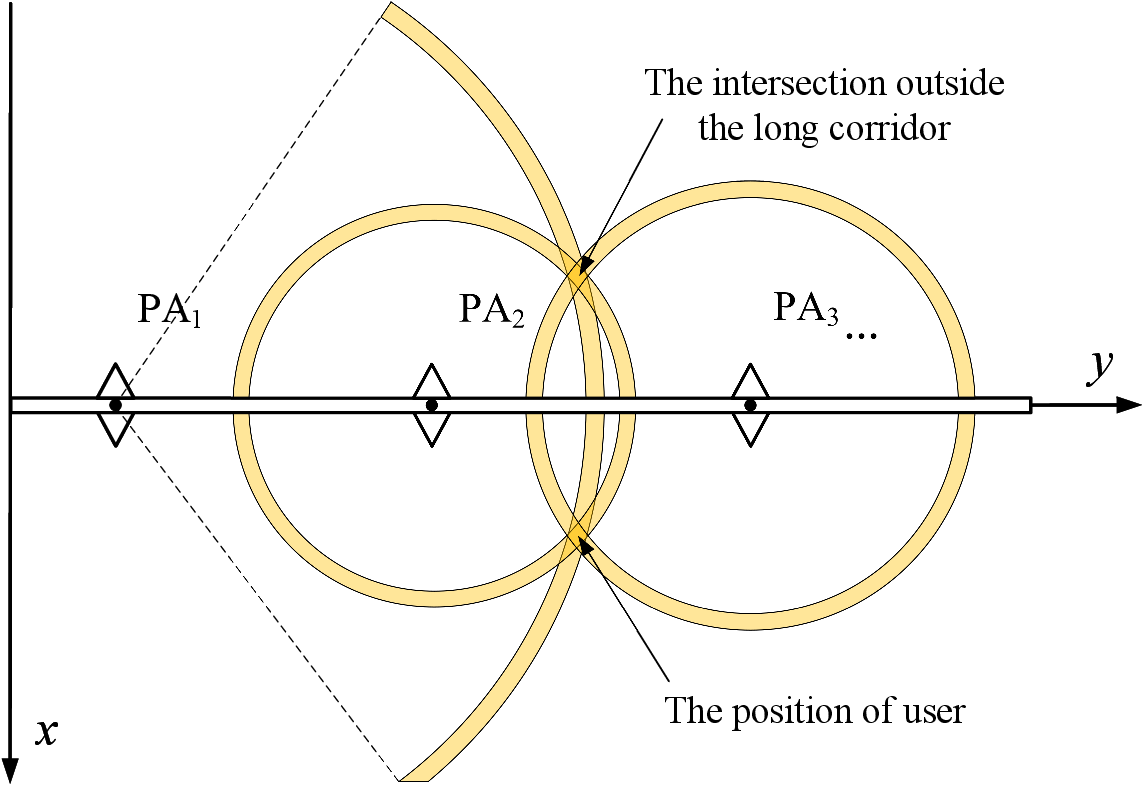}
\caption{Projection map of PASS-based navigation system.}
\label{Projection map}
\end{figure}

Geometrically, \eqref{navi_eq} defines a circle in the positioning plane $z=0$. Therefore, the solution of \eqref{system equation2} is the intersection point of these $N$ circles, which gives the estimated user position $\hat{\mathbf{u}}_k$.
In the signal waveguide scenarios shown in Fig.~\ref{system_model}, all of the PAs are collinear, so the centers of the circles defined by \eqref{navi_eq} lie on the same line. Consequently,~\eqref{system equation2} yields two feasible solutions. 
As shown in Fig.~\ref{Projection map}, the two solutions correspond to the two intersection points, which are symmetric with respect to the plane $x = 0$. 

The solution ambiguity makes it infeasible to directly solve \eqref{system equation2} by using the ordinary least squares (OLS) algorithm.
However, since the user is constrained within the long corridor with $x_{\mathrm{u},k} \in [0,D]$, the spurious solution outside the corridor can be uniquely identified and discarded. By using the spatial boundary constraint to eliminate the solution ambiguity, we define a new auxiliary variable $\hat{v}_k = \hat{x}_{\mathrm{u},k}^2 + \hat{y}_{\mathrm{u},k}^2$. Then, the original nonlinear navigation equations \eqref{system equation2} can be transformed into a closed-form linear system, given by 
\begin{equation}
\begin{cases}
- 2\hat{y}_{1k}\hat{y}_{\mathrm{u},k} + \hat{v}_k = \hat{b}_{1k} \\
- 2\hat{y}_{2k}\hat{y}_{\mathrm{u},k} + \hat{v}_k = \hat{b}_{2k} \\
\quad\quad\quad\quad\quad \vdots \\
- 2\hat{y}_{Nk}\hat{y}_{\mathrm{u},k} + \hat{v}_k = \hat{b}_{nk}
\end{cases},
\label{system equation3}
\end{equation}
where $\hat{b}_{nk} = \hat{d}_{nk}^2 - \hat{y}_{nk}^2 - h^2$. 
\begin{lemma}
In the PASS-based navigation approach, a linear closed-form system of navigation equations can be expressed as 
\begin{equation}
\hat{\mathbf{A}}_k \hat{\mathbf{x}}_k = \hat{\mathbf{b}}_k,
\label{martix}
\end{equation}
where
\begin{equation}
\hat{\mathbf{A}}_k = \begin{bmatrix}
-2\hat{y}_{1k} & 1 \ \\
-2\hat{y}_{2k} & 1 \ \\
\vdots & \vdots \ \\
-2\hat{y}_{Nk} & 1 \  
\end{bmatrix},  \nonumber
\end{equation}
\begin{equation}
\hat{\mathbf{x}}_k = \begin{bmatrix} \hat{y}_{\mathrm{u},k} & \hat{v}_k \ \end{bmatrix}^\mathrm{T}, \nonumber
\end{equation}
\begin{equation}
\hat{\mathbf{b}}_k = \begin{bmatrix} \hat{b}_{1k} & \hat{b}_{2k} & \cdots & \hat{b}_{nk} \end{bmatrix}^\mathrm{T}.\nonumber
\end{equation}
\begin{proof}
Rewriting~\eqref{system equation3} in matrix form, we can obtain \eqref{martix}, and the proof is complete.
\end{proof}
\end{lemma}
\begin{remark}
The navigation equations \eqref{martix} contains two unknowns, which means that a unique valid solution can be given when $N\geq2$. Therefore, the PASS-based navigation system requires at least two PAs to achieve user-side navigation in PASS. 
\end{remark}
\begin{remark}
When the number of PAs $N$ exceeds the number of unknowns, \eqref{system equation3} becomes an overdetermined equation, where the WLS algorithm solves $\hat{\mathbf{x}}_k$ by minimizing the sum of the weighted squared residuals. As the number of PAs increases, more valid independent measurements are incorporated into the solution of \eqref{martix}, so that the random errors in the measurement vector $\hat{\mathbf{b}}_k$ can effectively compensate for each other during the solution process, thus reducing the positioning error and improving the robustness of the positioning solution. 
\end{remark}
\begin{remark}
As the number of PAs increases, the random measurement errors in the observation vector $\hat{\mathbf{b}}_k$ in \eqref{martix} are statistically averaged throughout the solution procedure. Therefore, the susceptibility of the solution to individual noisy measurements is mitigated, thus improving the noise robustness of the navigation system. 
\end{remark}

In the proposed PASS-based navigation system, the effective channel gain and the AP-to-user propagation time via different PAs are distinct, leading to non-uniform reliability in PA positioning and pseudorange measurement.
To mitigate the performance degradation caused by low-reliability measurements and the nonlinear error amplification of the LWF, we develop an optimal WLS estimator with a normalized weight matrix, which is given in the following theorem.
\begin{theorem}\label{theorem_normalized_weighted_matrix}
In the PASS-based navigation approach, the optimal weight matrix for the WLS estimator is given by
\begin{equation}
\bar{\mathbf{\Omega}}_k = \frac{\mathrm{diag}\left( \bar\omega_{1k}, \bar\omega_{2k}, \cdots, \bar\omega_{Nk} \right)}{\sum_{n=1}^N \bar\omega_{nk}} ,
\label{eq:normalized_weight}
\end{equation}
where $\bar\omega_{nk}$ denotes the $n$-th optimal weight element, given by
\begin{equation}
\bar\omega_{nk} =\frac{\hat{P}_{nk}}{ \varepsilon_\mathrm{r} \hat{S}_{nk}^2 \hat{G}_{nk} \left( C_\mathrm{T} + C_\mathrm{P} \right) + {C_\mathrm{T} \hat{y}_{nk}^2} },
\label{eq:optimal_weight_element}
\end{equation}
in which $\hat{S}_{nk} = - W_0 ({\hat\zeta}_{nk} )  / ({ 1 + W_0 ({\hat\zeta}_{nk} ) + \varepsilon} )$ denotes the estimated Lambert sensitivity factor, $C_\mathrm{T} = c^2/(\pi^2 B^2)$ denotes the TOA measurement accuracy constant, $C_\mathrm{P} = 2 d_0^2/M$ denotes the power measurement accuracy constant, and $\hat{G}_{nk} = \hat{d}_{nk}^2 + \hat{y}_{nk}^2/\varepsilon_\mathrm{r}$ represents the estimated equivalent signal transmission distance. 
\begin{proof}
Please refer to Appendix B.
\end{proof}
\end{theorem}

By using \eqref{eq:optimal_weight_element}, the WLS solution to~\eqref{martix} is given by
\begin{equation}
\hat{\mathbf{x}}_k = \left( \hat{\mathbf{A}}^\mathrm{T}_k \bar{\mathbf{\Omega}}_k \hat{\mathbf{A}}_k \right)^{-1} \hat{\mathbf{A}}^\mathrm{T}_k \bar{\mathbf{\Omega}}_k \hat{\mathbf{b}}_k.
\label{WLS solution}
\end{equation}
There are two solutions for the $x$-coordinate of the user. We take the valid one satisfying $\hat{x}_{\mathrm{u},k} \in [0,D]$, which can be written by 
\begin{equation}
\hat{x}_{\mathrm{u},k} = \sqrt{\hat{v}_k - \hat{y}_{\mathrm{u},k}^2}.
\label{z-coordinate}
\end{equation}
Based on~\eqref{WLS solution} and~\eqref{z-coordinate}, the $k$-th user can estimate their position as $\hat{\mathbf{u}}_k = [\hat{x}_{\mathrm{u},k}, \hat{y}_{\mathrm{u},k}, 0]$. Then, the WLS-PAN algorithm can be given by \textbf{Algorithm~\ref{Algorithm 2}}. 
\begin{algorithm}[H]
\caption{WLS-PAN Algorithm} \label{Algorithm 2}
\begin{algorithmic}[1] 
\REQUIRE 
The waveguide geometry $\mathbf{G}_n$, transmission timestamp $t_n$, transmission power $P_n$, measured reception time $\hat{t}_{nk}$, measured reception power $\hat{P}_{nk}$, path-loss parameter $\eta$, breakpoint distance $d_0$, and waveguide relative permittivity ${\varepsilon_{\mathrm{r}}}$.
\ENSURE
The user position $\hat{\mathbf{u}}_k = [\hat{x}_{\mathrm{u},k}, \hat{y}_{\mathrm{u},k}, 0]$.
\STATE Obtain the estimated PA positions $\hat{\mathbf{w}}_{nk}=[0,\hat{y}_{nk},h]^{\mathrm{T}}$ and the estimated pseudorange $\hat{d}_{nk}$ by Algorithm~\ref{Algorithm 1}, $\forall n$.
\STATE Define $\hat{b}_{nk} = \hat{d}_{nk}^2 - \hat{y}_{nk}^2 - h^2$, $\forall n$. 
\STATE Construct the matrices $\hat{\mathbf{A}}_k$ and $\hat{\mathbf{b}}_k$ by~\eqref{martix}.
\STATE Construct the normalized weight matrix $\bar{\mathbf{\Omega}}_k$ by~\eqref{eq:normalized_weight}. 
\STATE Compute the solution $\hat{\mathbf{x}}_k = \begin{bmatrix} \hat{y}_{\mathrm{u},k} , \hat{v}_k \end{bmatrix}^\mathrm{T}$ by \eqref{WLS solution}. 
\STATE Calculate $\hat{x}_{\mathrm{u},k} = \sqrt{\hat{v}_k - \hat{y}_{\mathrm{u},k}^2}$.
\STATE \textbf{return} $\hat{\mathbf{u}}_k = [\hat{x}_{\mathrm{u},k}, \hat{y}_{\mathrm{u},k}, 0]$.
\end{algorithmic}
\end{algorithm}

\subsection{PA-derived Position Dilution of Precision Analysis}
In this subsection, we evaluate the theoretical accuracy bound of the proposed WLS-PAN algorithm by deriving a PA-PDOP metric. The PA-PDOP reflects the combined effects of the geometric distribution of PAs, the nonlinear error amplification introduced by the LWF, and the quality weighting of the measurements.

Based on the WLS solution given in (37), the error covariance matrix of the estimated intermediate vector is first derived as
\begin{equation}
\mathbf{C}_k = \mathrm{Cov}\left( \hat{\mathbf{x}}_k - \mathbf{x}_k \right) 
= \left( \hat{\mathbf{A}}_k^\mathrm{T} \bar{\mathbf{\Omega}}_k \hat{\mathbf{A}}_k \right)^{-1},
\end{equation}
where $\mathbf{x}_k = \begin{bmatrix} y_{\mathrm{u},k}, v_k \end{bmatrix}^\mathrm{T}$ and $v_k = x_{\mathrm{u},k}^2 + y_{\mathrm{u},k}^2$.

Based on the first-order error propagation law, the error covariance matrix of the user position $\mathbf{u}_k$ is derived as \cite{Wang2020_PDOP}
\begin{equation}
\mathbf{C}_{\mathrm{u},k} = \mathbf{J}_k \mathbf{C}_k \mathbf{J}_k^\mathrm{T},
\end{equation}
where $\mathbf{J}_k$ is the Jacobian matrix that binds the intermediate vector error to the user position error, given by 
\begin{equation}
\mathbf{J}_k = \begin{bmatrix}
-\dfrac{\hat{y}_{\mathrm{u},k}}{\hat{x}_{\mathrm{u},k}} & \dfrac{1}{2\hat{x}_{\mathrm{u},k}} \\
1 & 0
\end{bmatrix}.
\end{equation}

The error variances of the $x$-axis and the $y$-axis of the estimated position for the $k$-th user are given by
\begin{equation}
\sigma_{\mathrm{ux},k}^2 = \left[ \mathbf{C}_{\mathrm{u},k} \right]_{11},
\end{equation}
\begin{equation}
\sigma_{\mathrm{uy},k}^2 = \left[ \mathbf{C}_{\mathrm{u},k} \right]_{22},
\end{equation}
respectively, where $\left[ \mathbf{C}_{\mathrm{u},k} \right]_{ii}$ denotes the $i$-th diagonal element of matrix $\mathbf{C}_{\mathrm{u},k}$.

Then, the PA-PDOP can be calculate as \cite{Teunissen2017_PDOP}
\begin{equation}
\gamma_k = \sqrt{\mathrm{tr}\left( \mathbf{C}_{\mathrm{u},k} \right)} = \sqrt{\sigma_{\mathrm{ux},k}^2 + \sigma_{\mathrm{uy},k}^2}.
\end{equation}

Different from the conventional geometric-only DOP metric, the proposed PA-PDOP fully incorporates the optimal weight matrix, which inherently suppresses the nonlinear error amplification near the LWF's branch point and down-weights low-reliability measurements. 
\begin{remark}
A smaller PA-PDOP value indicates a more favorable PA deployment geometry, weaker nonlinear error amplification, higher overall measurement quality, and thus a tighter theoretical accuracy bound for the user position estimate.
\end{remark}

\section{SIMULATION RESULTS}

In this section, numerical results are provided for the performance evaluation of the proposed LWF-PAP algorithm and the benchmark WLS-PAN algorithm in the PASS-based navigation system. The system is deployed in a typical indoor corridor with dimensions $x \in [0,10]$ m, $y \in [0,12]$ m, and $z \in [0,3]$ m. A single waveguide of length $L=12$ m is mounted along the ceiling boundary at a height of $h=3$ m, with $N=8$ PAs uniformly spaced along its entire length. 
To evaluate the impact of PA layout on navigation performance, two typical deployment schemes are considered: uniform PA deployment and random PA deployment.
The relative permittivity of the waveguide medium is $\varepsilon_{\mathrm{r}} = 2.08$ and the dielectric dissipation factor is $\tan\delta = 4\times 10^{-4}$~\cite{PTFE_Jha_Singh_2013}. 
The breakpoint distance under default parameters is approximately $d_0=15.9$ m. 
The system operates at a carrier frequency of $f_\mathrm{c}=15$ GHz with a signal bandwidth of $B=20$ MHz and a transmit power of $P_n=10$ W, for $n=1,2,\dots,N$. 
The power of the AWGN is $\sigma^2 = -174+10\lg \left(B\right)$ dBm \cite{Johnson1928_ThermalNoise}. 
The variance of the signal power measurement error is $\sigma^2_{\mathrm{p},nk} = {2\bar{P}_{nk}\sigma^2}/M$~\cite{Urkowitz1967}, where $M=128$ is the number of samples used for the power estimation. 
The variance of the TOA measurement error is $\sigma^2_{\mathrm{t},nk} = {3\sigma^2}/({2\pi^2 B^2\bar{P}_{nk}})$ \cite{wilding2018accuracy, Zhang2022_TOA_CRLB}.

\subsection{Performance of LWF-PAP Algorithm}
\begin{figure}[!t]
\centering
\includegraphics[width =3.0in]{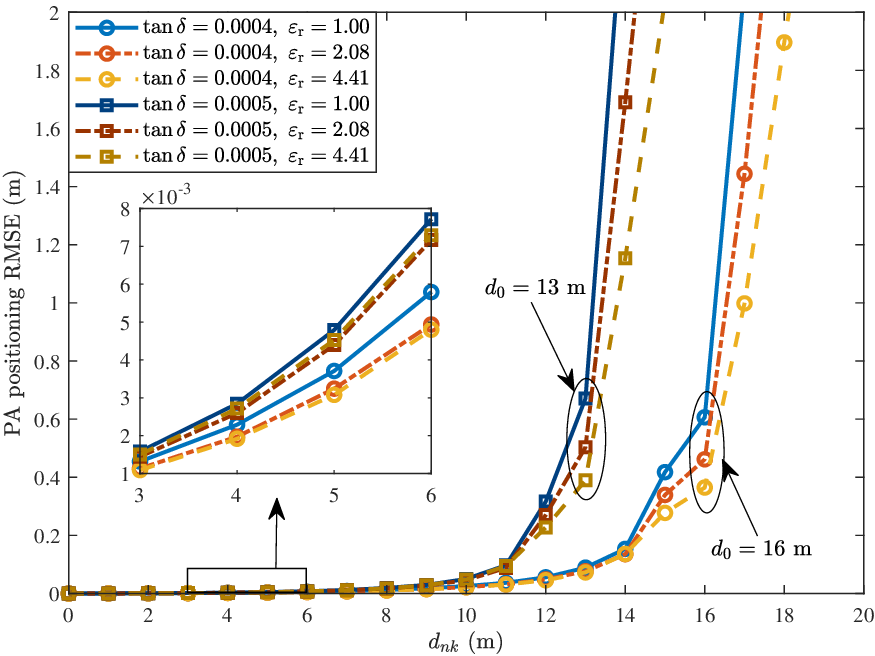}
\caption{PA positioning RMSE versus PA-user distance $d_{nk}$ for varying waveguide dielectric dissipation factor $\tan\delta$ and relative permittivity $\varepsilon_{\mathrm{r}}$.} 
\label{fig1_PAvsdnk}
\end{figure}

This subsection evaluates the positioning accuracy of individual PAs and the pseudorange estimation accuracy. 
Fig. \ref{fig1_PAvsdnk} shows the PA positioning RMSE as a function of the PA-user distance for different waveguide dielectric dissipation factors and relative permittivities. The PA positioning RMSE increases with the PA-user distance, and exhibits a sharp nonlinear growth after exceeding the breakpoint distance. The phenomenon arises because that measurement errors are dominated by additive white Gaussian noise within the breakpoint distance, while the nonlinear amplification effect of the LWF becomes the dominant error source as the distance approaches and exceeds the breakpoint. A larger dissipation factor leads to a smaller breakpoint distance, thus triggering the rapid error growth at a shorter PA-user distance. The influence of relative permittivity on positioning accuracy is not significant before the breakpoint, whereas after the breakpoint, the positioning error increases faster with smaller relative permittivity for a given dissipation factor.
Accordingly, adopting waveguide materials with low dissipation factor can effectively extend the reliable working distance and restrain sharp rise of positioning error.

\begin{figure}[t!]
\centering
\includegraphics[width =3.0in]{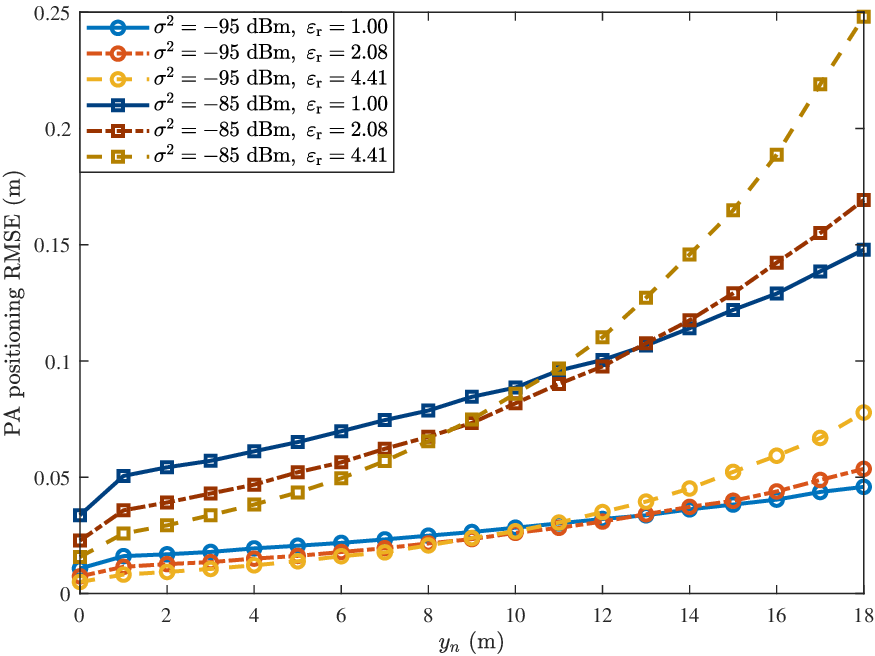}
\caption{PA positioning RMSE versus PA-AP distance $y_{n}$ for varying noise variance $\sigma^2$ and relative permittivity $\varepsilon_{\mathrm{r}}$.} 
\label{fig2_PAvsynk}
\end{figure}

Fig. \ref{fig2_PAvsynk} shows the PA positioning RMSE as a function of the PA-AP distance for different noise variances and relative permittivities. The PA positioning RMSE increases monotonically with the PA-AP distance due to the cumulative attenuation of signals propagating along the waveguide. 
An increase in noise variance consistently impairs the PA positioning accuracy.
As the PA-AP distance increases, the resulting decrease in the received signal power further worsens the PA positioning performance.
The influence of relative permittivity on positioning accuracy exhibits a non-monotonic crossover behavior. At small PA-AP distances, a larger relative permittivity yields lower RMSE because that the reduced group velocity in the waveguide decreases the sensitivity of the position estimate to TOA measurement errors. At larger distances, conversely, a smaller relative permittivity leads to better positioning performance since it results in a lower waveguide attenuation coefficient, which maintains higher received signal power and reduces measurement noise. 
Accordingly, high permittivity media fit short-distance transmission scenarios, while low permittivity materials deliver better performance for long-range waveguide propagation, and noise suppression helps stabilize positioning accuracy in all cases.

\begin{figure}[t!]
\centering
\includegraphics[width =3.0in]{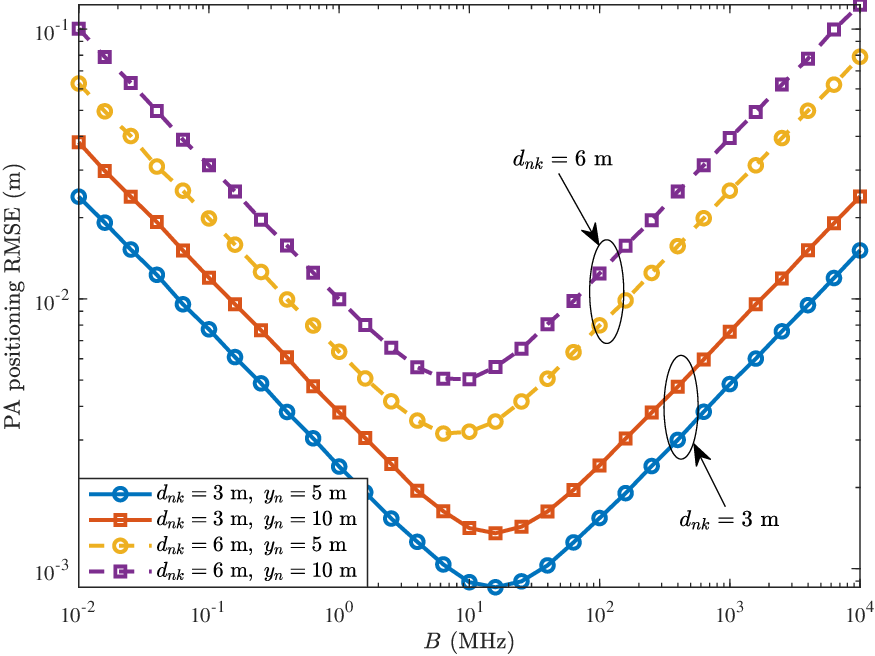}
\caption{PA positioning RMSE versus navigation bandwidth $B$ for varying PA-user distance $d_{nk}$ and PA-AP distance $y_{n}$.} 
\label{fig5_PAvsB}
\end{figure}

Fig. \ref{fig5_PAvsB} shows the PA positioning RMSE as a function of the signal bandwidth for different combinations of PA-user distance and PA-AP distance. The PA positioning RMSE exhibits a distinct U-shaped relationship with the signal bandwidth, first decreasing rapidly and then increasing after reaching an optimal value. The phenomenon arises because that the improvement in TOA measurement accuracy dominates at low bandwidths, while the increase in total noise power proportional to bandwidth becomes the dominant factor at high bandwidths. A larger PA-user or PA-AP distance leads to a higher overall positioning RMSE across all bandwidth values, as the received signal power decreases with increasing distance. The optimal bandwidth that minimizes the positioning error also shifts to lower values as the user-AP distance increases, due to the reduced SNR at longer ranges.
Accordingly, both excessively low and high bandwidth will degrade positioning performance, and moderate bandwidth value contributes to achieving the best positioning accuracy.

\begin{figure}[t!]
\centering
\includegraphics[width =3.0in]{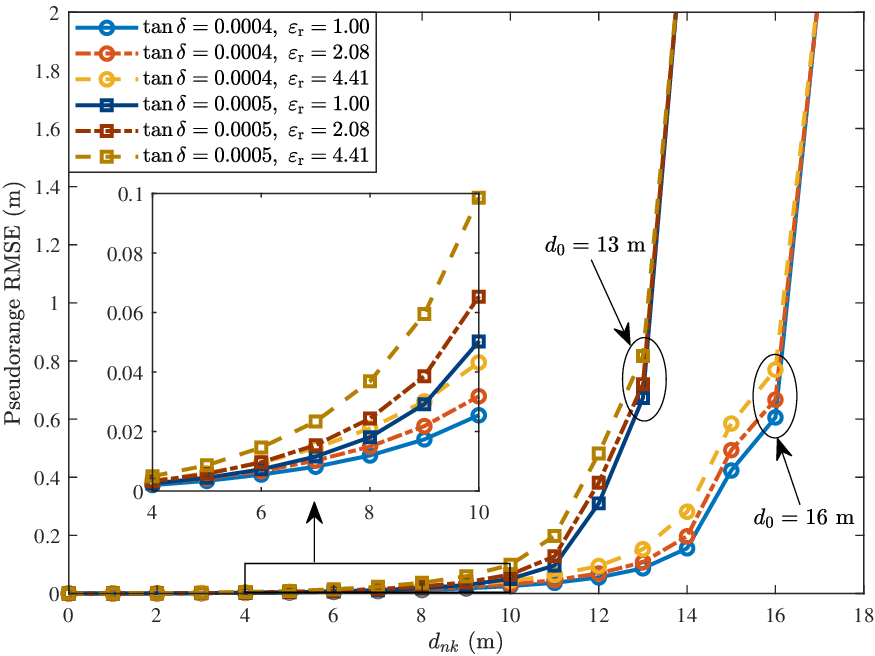}
\caption{Pseudorange RMSE versus PA-user distance $d_{nk}$ for varying waveguide dielectric dissipation factor $\tan\delta$ and relative permittivity $\varepsilon_{\mathrm{r}}$.} 
\label{fig3_Pseudovsdnk}
\end{figure}

Fig. \ref{fig3_Pseudovsdnk} shows the pseudorange RMSE as a function of the PA-user distance for different waveguide dielectric dissipation factors and relative permittivities. The pseudorange RMSE increases with the PA-user distance, and exhibits a sharp growth after exceeding the breakpoint distance, which is consistent with the trend observed in the PA positioning error. The similarity arises because that both the pseudorange estimate and the PA position estimate are derived from the same TOA and received power measurements through the LWF. A larger dissipation factor leads to a smaller breakpoint distance, thus triggering the rapid error growth at a shorter PA-user distance. A larger relative permittivity leads to a slightly higher pseudorange RMSE, but the influence of relative permittivity is negligible compared to the effect of the dissipation factor.
Accordingly, the dielectric dissipation factor plays a dominant role in overall system performance, while relative permittivity exerts barely noticeable impact on pseudorange measurement.

\subsection{Performance of WLS-PAN Algorithm}
\begin{figure}[t!]
\centering
\includegraphics[width =3.5in]{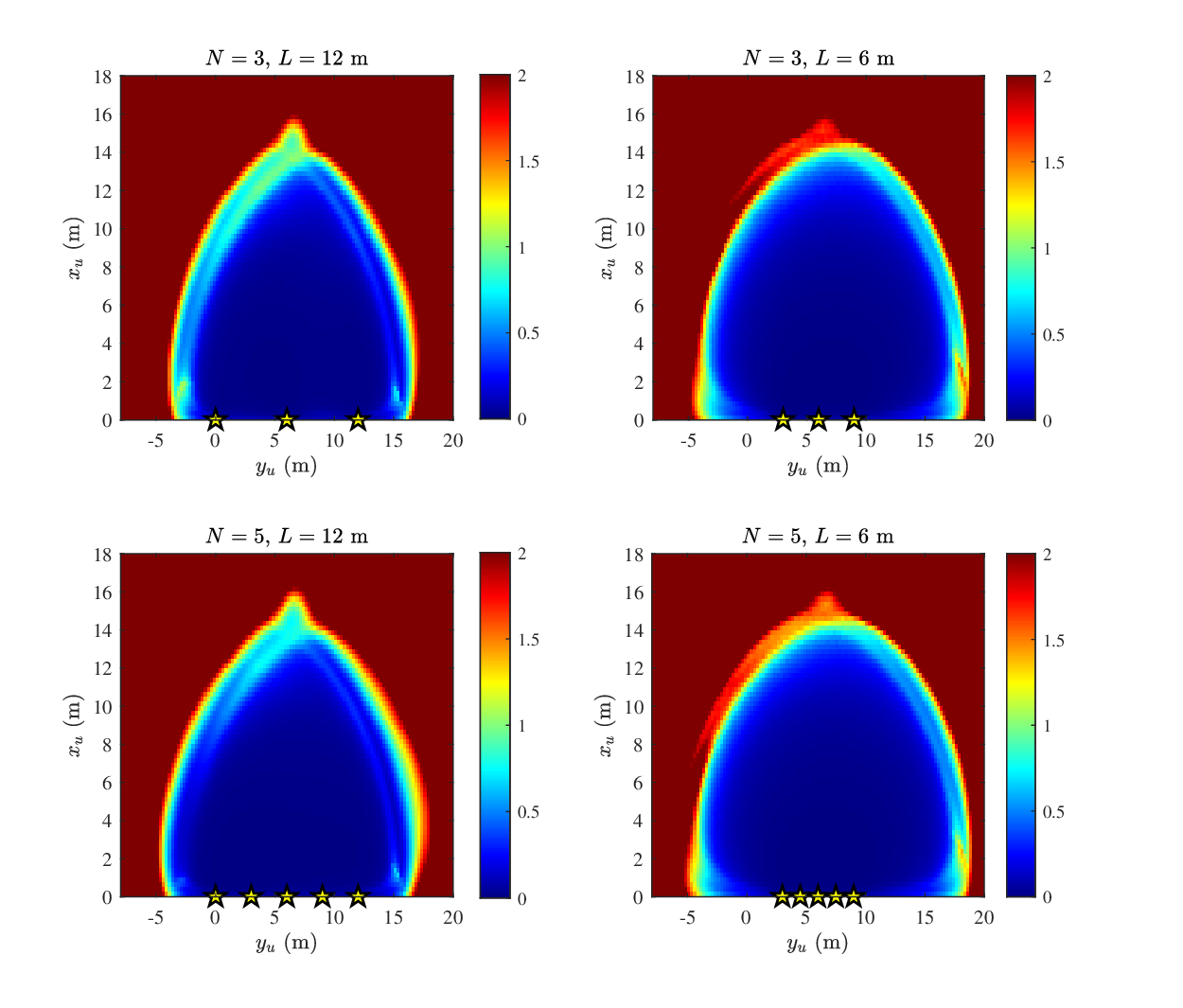}
\caption{User positioning RMSE versus user position $[x_{\mathrm{u},k},y_{\mathrm{u},k}]$ for varying number of PAs $N$ and waveguide lengths $L$.} 
\label{Heatmap}
\end{figure}
This subsection investigates the user-side navigation performance of the proposed weighted least squares-based algorithm. 
Fig.~\ref{Heatmap} illustrates the spatial distribution of user positioning RMSE with varying number of PAs and waveguide lengths. 
The positioning error stays at a low level within the breakpoint distance, and rises sharply once the user moves beyond this range as measurement reliability degrades significantly. 
Under the same waveguide length, increasing the number of PAs consistently enhances overall positioning performance. More PA nodes enrich the angular diversity of received measurements and improve geometric observability, which suppresses positioning errors in edge areas and further enlarges the effective high-precision coverage.


\begin{figure}[t!]
\centering
\includegraphics[width =3.0in]{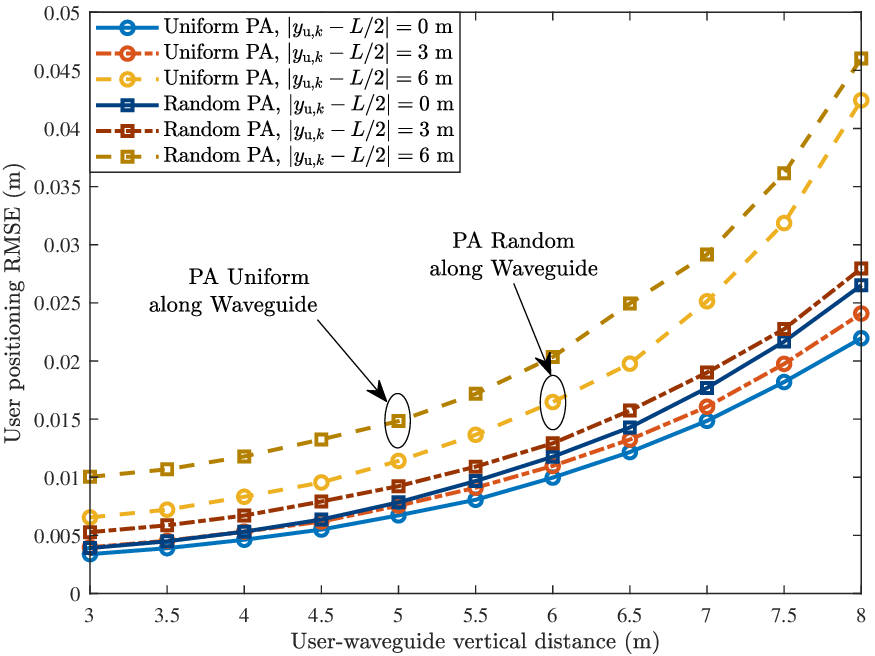}
\caption{User positioning RMSE versus user-waveguide vertical distance $d_{\mathrm{w},k}$ for varying PA deployment schemes and user lateral offsets from the waveguide center $|y_{\mathrm{u},k}-L/2|$.} 
\label{fig7_UEvsdu}
\end{figure}
Fig. \ref{fig7_UEvsdu} shows the user positioning RMSE as a function of the user-waveguide vertical distance $d_{\mathrm{w},k} = \sqrt{x_{\mathrm{u},k}^2 + h^2}$ for different PA deployment schemes and user lateral offsets from the waveguide center $|y_{\mathrm{u},k}-L/2|$. The user positioning RMSE increases monotonically with the user-waveguide vertical distance, as the distance between the user and all PAs increases, leading to lower received signal power and higher measurement errors. Uniform PA deployment consistently outperforms random deployment across all distances and lateral offsets, as the uniform distribution provides better geometric coverage and more balanced measurement weights. A larger user lateral offset from the waveguide center results in higher positioning error, which is attributed to the degraded geometric dilution of precision when the user is located near the ends of the waveguide.
Accordingly, uniform PA deployment outperforms random deployment, and users near the waveguide center obtain superior navigation performance compared with edge users.

\begin{figure}[t!]
\centering
\includegraphics[width =3.0in]{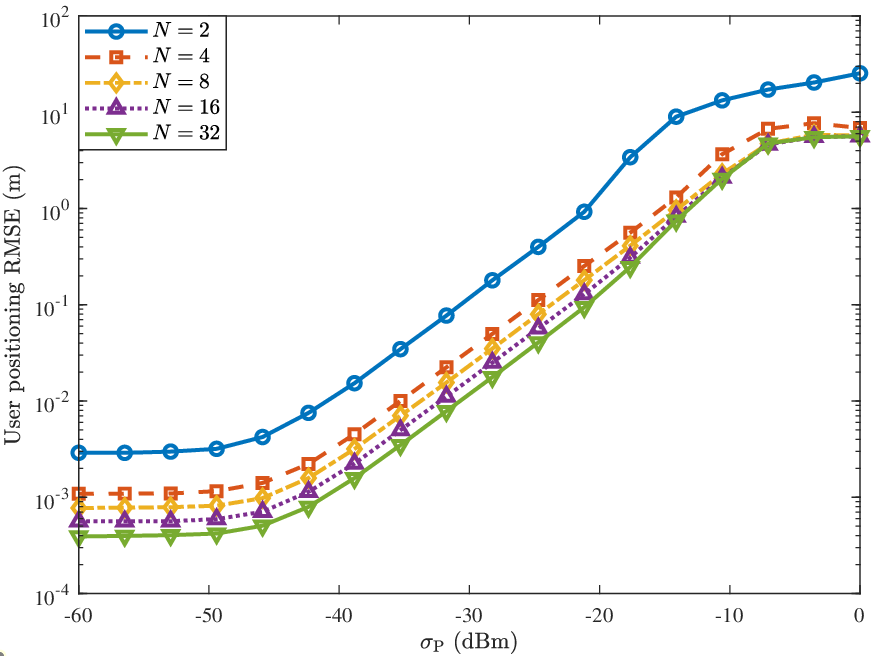}
\caption{User positioning RMSE versus standard deviation of noise power $\sigma_{\mathrm{P}}$ for varying number of PAs $N$.} 
\label{fig8_UEvsSigmaP}
\end{figure}

Fig. \ref{fig8_UEvsSigmaP} shows the user positioning RMSE as a function of the standard deviation of power measurement noise for different numbers of PAs. The user positioning RMSE exhibits a three-stage behavior with respect to the power measurement noise. At low noise levels, the RMSE saturates at a lower bound and does not decrease further as the power measurement noise reduces, which is because that the TOA measurement error, rather than power measurement noise, becomes the dominant error source. In the mid-range of the power measurement noise, the RMSE increases monotonically, and this performance degradation becomes more pronounced at higher noise levels. At sufficiently high noise levels, the RMSE saturates again and does not increase further with the power measurement noise, indicating that the power measurement is no longer reliable and the positioning performance is limited by other system constraints. Increasing the number of PAs effectively mitigates the impact of power measurement noise in the mid-range region, as the fusion of multiple independent measurements reduces the overall error variance, leading to a lower RMSE for a given power measurement noise. However, when the power measurement noise is large, the improvement brought by increasing the number of PAs diminishes, and the RMSE approaches an error floor, which is because that the high noise level corrupts the power measurements such that additional PAs no longer provide meaningful information to improve the position estimate. 
Accordingly, the multi-PA fusion architecture significantly improves the robustness of the navigation system against moderate power measurement errors, but its effectiveness is limited at extremely high noise levels.

\begin{figure}[t!]
\centering
\includegraphics[width =3.0in]{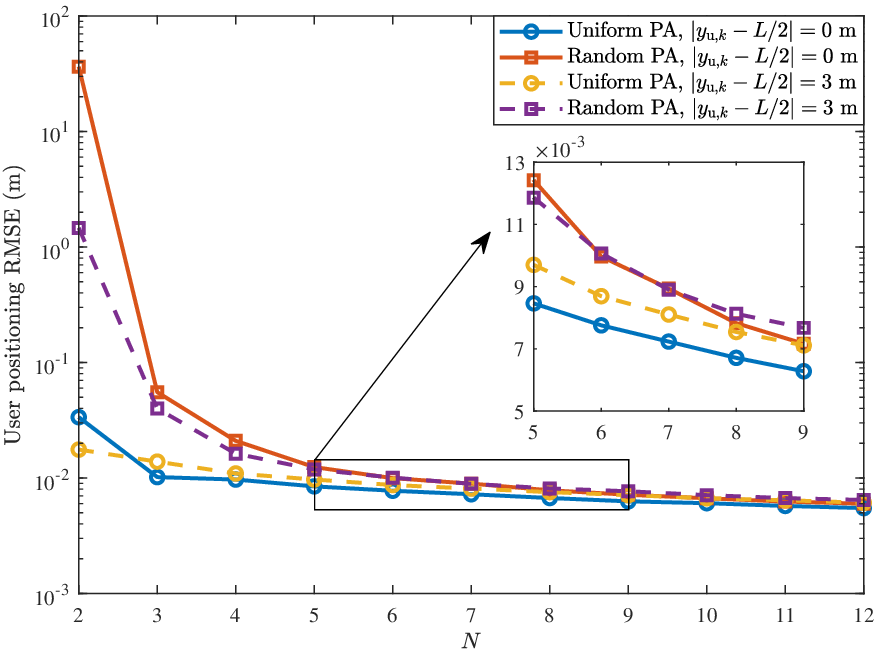}
\caption{User positioning RMSE versus number of PAs $N$ for varying PA deployment schemes and user lateral offsets from the waveguide center $|y_{\mathrm{u},k}-L/2|$.} 
\label{fig12_UEvsN}
\end{figure}
Fig. \ref{fig12_UEvsN} shows the user positioning RMSE as a function of the number of PAs for different PA deployment schemes and user lateral offsets from the waveguide center. The user positioning RMSE decreases remarkably as the number of PAs increases at low PAs deployment scale, and the declining trend becomes gentle after the quantity reaches a certain level. The phenomenon arises because that increasing the number of PAs provides more independent measurement information and improves the system geometric diversity, but the performance improvement exhibits a diminishing return as the measurement redundancy approaches saturation. Uniform PA deployment consistently outperforms random deployment, and a smaller user lateral offset from the waveguide center yields better positioning accuracy. 
Accordingly, it is unnecessary to blindly increase the number of PAs in practical deployment, and rational PA configuration together with optimized layout can achieve favorable navigation performance.

\subsection{PA-PDOP Analysis}
\begin{figure}[t!]
\centering
\includegraphics[width =3.0in]{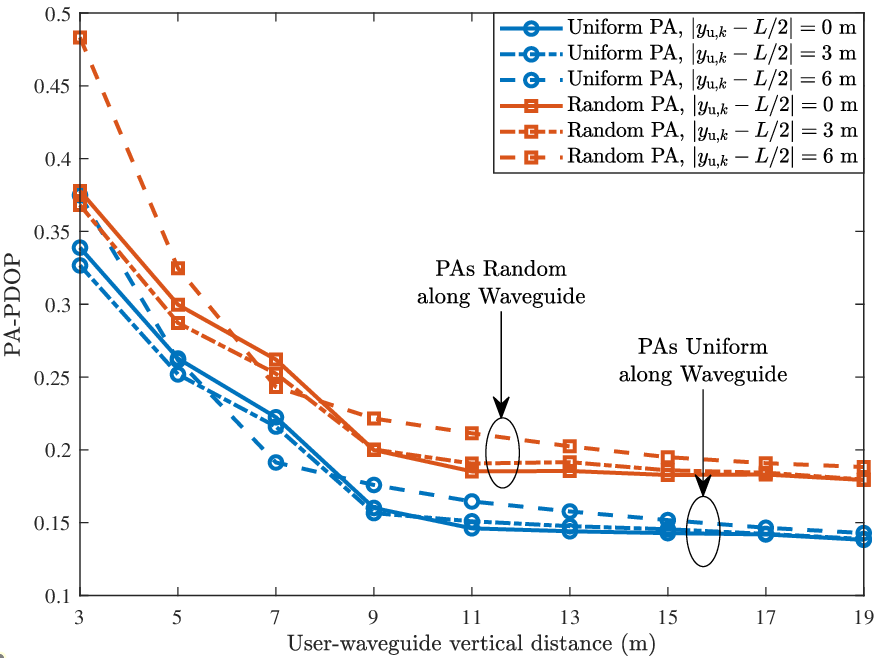}
\caption{PA-PDOP versus user-waveguide vertical distance $d_{\mathrm{w},k}$ for varying waveguide dielectric dissipation factor $\tan\delta$ and user lateral offsets from the waveguide center $|y_{\mathrm{u},k} - L/2|$.} 
\label{PAPDOPvsdUW}
\end{figure}

This subsection analyzes the PA-PDOP performance under different system parameters. 
Fig. \ref{PAPDOPvsdUW} illustrates the PA-PDOP variation with respect to the user-waveguide vertical distance for different PA deployment schemes and user lateral offsets from the waveguide center. The PA-PDOP decreases rapidly as the user-waveguide vertical distance increases initially, and gradually approaches a constant value with further increase in distance. 
The underlying cause of the variation lies in the inherent linear topology constraint of the single-waveguide PASS system, where all PAs are distributed along a one-dimensional straight line of the waveguide. 
Specifically, when the user is in close proximity to the waveguide, the one-dimensional PA arrangement results in extremely poor geometric observability for the x-coordinate while maintaining excellent observability for the y-coordinate. As the user moves away from the waveguide, the angular separation between the user and different PAs increases significantly, which substantially improves the x-coordinate observability and reduces its estimation error, albeit at the cost of a moderate degradation in y-coordinate observability. 
Since the x-coordinate error dominates the overall positioning error of the PASS-based navigation system, the PA-PDOP thus decreases continuously as the vertical user-waveguide distance increases.
For the PA deployment scheme, uniform PA deployment consistently yields lower PA-PDOP values compared to random deployment across all considered distances and lateral offsets. The superiority stems from the fact that uniform distribution provides more balanced geometric coverage and more symmetrical measurement geometry, which effectively reduces the geometric dilution of precision.
Furthermore, the impact of user lateral offset on PA-PDOP exhibits a distinct distance-dependent characteristic. When the user is very close to the waveguide, a moderate lateral offset from the waveguide center creates a larger angular difference between the LoS vectors to adjacent PAs, resulting in a better geometric configuration and thus lower PA-PDOP. 
When the user is far from the waveguide, the optimal geometric configuration is achieved when the user is directly aligned with the waveguide center, maximizing the overall angular spread of PA measurements relative to the user.


\section{Conclusions}
In this paper, we proposed a PASS-based user-side navigation framework, in which an AP broadcast downlink signals through multiple PAs along a single waveguide. Each user independently estimated its own position without any prior knowledge of the PA positions. To handle the unknown PA positions and the coupled propagation components, a two-step approach was developed. We first developed the LWF-PAP algorithm, which yielded closed-form expressions for both the PA positions and the PA-user pseudoranges using TOA and received power measurements. Then, we developed the WLS-PAN algorithm, which optimally fused multiple PA measurements to compute the user position. We also introduced a PA-PDOP metric to characterize the theoretical accuracy bound.

Several promising directions remain for future research. One extension is to investigate multi-waveguide configurations to improve geometric diversity and further reduce positioning errors. Another direction is to resolve the solution ambiguity caused by the multi-branch nature of the LWF beyond the breakpoint distance. Furthermore, integrating the proposed navigation scheme with communication functionalities under the integrated navigation and communications (INAC) framework represents a promising direction for future work.



\numberwithin{equation}{section}
\section*{Appendix~A: Proof of Lemma~\ref{lemmacondition}} \label{Appendix:As}
\renewcommand{\theequation}{A.\arabic{equation}}
\setcounter{equation}{0}
This appendix provides the detailed derivation of~\eqref{condition}. By eliminating ${y}_{n}$ and isolate ${d}_{nk}$ from~\eqref{system equation1}, we have 
\begin{equation}
\ln({d}_{nk}) - \frac{\alpha}{\sqrt{\varepsilon_{\mathrm{r}}}} {d}_{nk} = \frac{\ln 10}{20} \bar{L}_{nk} |_{\mathrm{dB}} - \frac{\alpha c\bar{T}_{nk}}{\sqrt{\varepsilon_{\mathrm{r}}}} + \ln(\eta).
\label{eq:appendix_A1}
\end{equation}

Note that the right-hand side (RHS) of~\eqref{eq:appendix_A1} is a constant, and to simplify the subsequent derivation, we define 
\begin{equation}
\bar{\xi}_{nk} = \frac{\ln 10}{20} \bar{L}_{nk} |_{\mathrm{dB}} - \frac{\alpha c\bar{T}_{nk}}{\sqrt{\varepsilon_{\mathrm{r}}}} + \ln(\eta).
\label{eq:appendix_A2}
\end{equation}

Then,~\eqref{eq:appendix_A1} can be written as 
\begin{equation}
\ln({d}_{nk}) - \frac{\alpha}{\sqrt{\varepsilon_{\mathrm{r}}}} {d}_{nk} = \bar{\xi}_{nk}.
\label{eq:appendix_A3}
\end{equation}

It is evident that~\eqref{eq:appendix_A3} is a transcendental equation for ${d}_{nk}$, which cannot be solved directly via conventional algebraic methods. 
To address the transcendental equation, we resort to the LWF, which is defined as the inverse function of the transcendental function $f(t) = t e^t$ such that $t = W(z)$ if and only if $z = t e^t$ \cite{LambertWFunction1,LambertWFunction2}. We thus proceed to transform~\eqref{eq:appendix_A3} into the standard form of the LWF. By some algebraic manipulations, we have 
\begin{equation}
-\frac{\alpha}{\sqrt{\varepsilon_{\mathrm{r}}}} {d}_{nk} \exp\left(-\frac{\alpha}{\sqrt{\varepsilon_{\mathrm{r}}}} {d}_{nk}\right) = -\frac{\alpha e^{\bar{\xi}_{nk}}}{\sqrt{\varepsilon_{\mathrm{r}}}}, 
\label{eq:appendix_A4}
\end{equation}
which matches the standard LWF form on substituting $t = -\frac{\alpha d_{nk}}{\sqrt{\varepsilon_{\mathrm{r}}}}$. By the definition of the LWF, we can obtain 
\begin{equation}
-\frac{\alpha}{\sqrt{\varepsilon_{\mathrm{r}}}} {d}_{nk} = W\left(-\frac{\alpha e^{\bar{\xi}_{nk}}}{\sqrt{\varepsilon_{\mathrm{r}}}} \right), 
\label{eq:appendix_A5}
\end{equation}
where $W(\cdot)$ is the LWF. 

It is important to note that LWF has two distinct real branches, namely the principal branch $W_0(\cdot)$ and the negative branch $W_{-1}(\cdot)$, where $W_0(\cdot) \geq -1 $ and $W_{-1}(\cdot) < -1$ hold. Thus, \eqref{eq:appendix_A5} can be rewritten by 
\begin{equation}
\begin{split}
-\frac{\alpha}{\sqrt{\varepsilon_{\mathrm{r}}}} {d}_{nk} = 
\begin{cases}
W_0\left(-\dfrac{\alpha e^{\bar{\xi}_{nk}}}{\sqrt{\varepsilon_{\mathrm{r}}}} \right) \geq -1, \\
W_{-1}\left(-\dfrac{\alpha e^{\bar{\xi}_{nk}}}{\sqrt{\varepsilon_{\mathrm{r}}}} \right) < -1.
\end{cases}
\end{split}
\label{eq:appendix_dnk}
\end{equation}

For the proposed PASS-based navigation approach, the system is inherently characterized by short-range LoS transmission between PAs and users, which ensures that the left-hand side of \eqref{eq:appendix_A5} generally falls within the range greater than $-1$, matching the valid range of the LWF's principal branch $W_0(\cdot)$. 
Thus, to guarantee the existence and uniqueness of the solution, we take 
\begin{equation}
-\frac{\alpha}{\sqrt{\varepsilon_{\mathrm{r}}}} {d}_{nk} = W_0\left(-\frac{\alpha e^{\bar{\xi}_{nk}}}{\sqrt{\varepsilon_{\mathrm{r}}}} \right), 
\label{eq:appendix_A6}
\end{equation}

Then, the distance between the $n$-th PA and the $k$-th user $d_{nk}$ can admit a unique closed-form solution. By using \eqref{T_total}, the distance between the AP and the $n$-th PA $y_{n}$ can also admit a unique closed-form solution. 

Based on \eqref{eq:appendix_dnk} and \eqref{eq:appendix_A6}, we have 
\begin{equation}
-\frac{\alpha}{\sqrt{\varepsilon_{\mathrm{r}}}} {d}_{nk} \geq -1. 
\label{eq:appendix_A8}
\end{equation}

By some algebraic manipulations, we can obtain~\eqref{condition}, and the proof is complete.

\numberwithin{equation}{section}
\section*{Appendix~B: Proof of Theorem~\ref{theorem_normalized_weighted_matrix}} \label{Appendix:Bs}
\renewcommand{\theequation}{B.\arabic{equation}}
\setcounter{equation}{0}
This appendix provides the complete derivation of the normalized optimal weight matrix, including the variance of the navigation observation error and the construction of the optimal weight matrix.

First, we derive the variance of the $n$-th navigation observation error $\sigma_{\mathrm{b},nk}^2$. The navigation observation is defined as $\hat{b}_{nk} = \hat{d}_{nk}^2 - \hat{y}_{nk}^2$, where $\hat{d}_{nk}$ is the estimated pseudorange and $\hat{y}_{nk}$ is the estimated PA position. The true value of the observation is $b_{nk} = d_{nk}^2 - y_n^2$, where $d_{nk}$ and $y_n$ are the true pseudorange and true PA position, respectively. The observation error is denoted as 
$\Delta b_{nk} = \hat{b}_{nk} - b_{nk}$.

By using \eqref{eq:dnk_variance} and \eqref{eq:ynk_variance}, the variance of the observation error is derived as
\begin{equation}
\begin{split}
\sigma_{\mathrm{b},nk}^2 &= \mathrm{Var}(\Delta b_{nk}) \\
&= \left( \frac{\partial b_{nk}}{\partial d_{nk}} \right)^2 \sigma_{\mathrm{d},nk}^2 + \left( \frac{\partial b_{nk}}{\partial y_n} \right)^2 \sigma_{\mathrm{y},nk}^2 \\
&= \left[ 4 c^2 S_{nk}^2 \zeta_{nk}^2 \left( d_{nk}^2 + \frac{y_n^2}{\varepsilon_\mathrm{r}} \right) + \frac{4 c^2 y_n^2}{\varepsilon_\mathrm{r}} \right] \sigma_{\mathrm{t},nk}^2  \\
&+ \left[ \frac{d_0^2 S_{nk}^2 \zeta_{nk}^2}{P_{nk}^2} \left( d_{nk}^2 + \frac{y_n^2}{\varepsilon_\mathrm{r}} \right) \right] \sigma_{\mathrm{p}}^2.
\end{split}
\label{eq:sigma_bnk}
\end{equation}

In addition, the variance of the received power measurement error $\hat{\sigma}_\mathrm{p}^2$ is derived from the statistical characteristics of energy detection-based power measurement in AWGN environments. For a received signal with $M$ independent sampling points, the variance of the power measurement error is given by
\begin{equation}
{\sigma}_\mathrm{p}^2 = \frac{2 {P}_{nk} \sigma^2}{M}.
\label{eq:power_variance}
\end{equation}

The variance of the TOA measurement error ${\sigma}^2_\mathrm{t}$ can be determined by the CRLB of time delay estimation in AWGN environments, which is given by
\begin{equation}
{\sigma}^2_\mathrm{t} = \frac{\sigma^2}{4\pi^2 B^2 \hat{P}_{nk}}.
\label{eq:toa_crlb}
\end{equation}

Based on \eqref{eq:sigma_bnk}, \eqref{eq:power_variance} and \eqref{eq:toa_crlb}, the $k$-th user can estimate the variance of the observation error as 
\begin{equation}
\hat\sigma_{\mathrm{b},nk}^2 = \frac{\hat{\sigma}^2}{\hat{P}_{nk}} \left[ \hat{S}_{nk}^2 \hat{G}_{nk} \left( C_\mathrm{T} + C_\mathrm{P} \right) + \frac{C_\mathrm{T} \hat{y}_{nk}^2}{\varepsilon_\mathrm{r}} \right],
\label{eq:sigma_bnk_simplified}
\end{equation}
where $C_\mathrm{T} = c^2/(\pi^2 B^2)$ denotes the TOA measurement accuracy constant, $C_\mathrm{P} = 2 d_0^2/M$ denotes the power measurement accuracy constant, $\hat{G}_{nk} = \hat{d}_{nk}^2 + \hat{y}_{nk}^2/\varepsilon_\mathrm{r}$ represents the geometric term related to the PA position and pseudorange estimate, and $\hat{S}_{nk}$ denotes the estimated Lambert sensitivity factor, given by
\begin{equation}
\begin{split}
\hat{S}_{nk} = - \frac{ W_0 ({\hat\zeta}_{nk} )}{1 + W_0 ({\hat\zeta}_{nk} )  + \varepsilon}.
\end{split}
\label{eq:hat_lambert_sensitivity}
\end{equation}

For the linear overdetermined system $\hat{\mathbf{A}}_k \hat{\mathbf{x}}_k = \hat{\mathbf{b}}_k$, the optimal weight matrix that achieves the minimum variance unbiased estimation is the inverse of the observation error covariance matrix. Since the navigation observations from different PAs are independent, the observation error covariance matrix is a diagonal matrix, and thus the optimal weight matrix is also a diagonal matrix.
Based on the Gauss-Markov theorem, the optimal weight matrix can be given by \cite{Kay1993_Estimation}
\begin{equation}
\hat{\mathbf{\Omega}}_k = \mathrm{diag}\left( \hat\omega_{1k}, \hat\omega_{2k}, \cdots, \hat\omega_{Nk} \right),
\label{eq:optimal_weight_matrix}
\end{equation}
where the $n$-th diagonal element is the optimal weight for the $n$-th navigation observation, defined as
\begin{equation}
\begin{split}
\hat\omega_{nk} &= \frac{1}{\hat\sigma_{\mathrm{b},nk}^2} \\
&=\frac{\varepsilon_\mathrm{r}\hat{P}_{nk}}{ {\hat{\sigma}^2} \left[ \varepsilon_\mathrm{r} \hat{S}_{nk}^2 \hat{G}_{nk} \left( C_\mathrm{T} + C_\mathrm{P} \right) + {C_\mathrm{T} \hat{y}_{nk}^2} \right] }.
\end{split}
\label{eq:optimal_weight_element_app}
\end{equation}

To avoid numerical singularity and improve the stability of the WLS solver, we normalize the weight matrix $\hat{\boldsymbol{\Omega}}_k$ by trace normalization, defining the normalized weight matrix as
\begin{equation}
\bar{\mathbf{\Omega}}_k = \frac{\hat{\mathbf{\Omega}}_k}{\mathrm{tr}\left( \hat{\mathbf{\Omega}}_k \right)}, 
\label{eq:normalization_rule}
\end{equation}
where $\mathrm{tr}\left( \cdot \right)$ denotes the matrix trace operator. 
For a diagonal matrix, the trace is equal to the sum of all its diagonal elements. Therefore, the trace of the weight matrix can be given by
\begin{equation}
\mathrm{tr}\left( \hat{\mathbf{\Omega}}_k \right) = \sum_{n=1}^N \hat\omega_{nk}.
\label{eq:weight_trace}
\end{equation}



By substituting \eqref{eq:optimal_weight_matrix} and \eqref{eq:weight_trace} into \eqref{eq:normalization_rule}, we can obtain~\eqref{eq:normalized_weight}, and the proof is complete.

\bibliographystyle{IEEEtran}
\bibliography{IEEEabrv,NOMA_RIS_INAC}

\end{document}